# IMPROVING THE COMMUNICATION FOR CHILDREN WITH SPEECH DISORDERS USING THE SMART TOYS

Amr Mohsen Jadi

Department of Computer Science and Information, University of Hail, Saudi Arabia

## ABSTRACT

*An attempt is made to develop a smart toy to help the children suffering with communication disorders. The children suffering with such disorders need additional attention and guidance to understand different types of social events and life activities. Various issues and features of the children with speech disorders are identified in this study and based on the inputs from the study, a working architecture is proposed with suitable policies. A prediction module with a checker component is designed in this work to produce alerts in at the time of abnormal behaviour of the child with communication disorder. The model is designed very sensitively to the behaviour of the child for a particular voice tone, based on which the smart toy will change to tones automatically. Such an arrangement proved to be helpful for the children to improve the communication with other due to the inclusion of continuous training for the smart toy from the prediction module.*

## 1. INTRODUCTION

In the recent times, kids with speech disorders found to be facing a serious discrimination within the society at schools, public places, and even within their families too. There are varieties of reasons for such disorders and kids behaviour which comes out of neurological issues at the time of birth. Cerebral palsy is the general term used for such neurological conditions, which in turn influence the movement of kids, their muscle tone, tremors, weak muscles, stiff muscles, etc [1]. Some of the causes for such disorders include infant infections, brain injuries, and restricted flow of blood to the brain (i.e. due to blood clot, accidents, strokes and bleeding in the brain. Children with cerebral palsy are provided with a special training in many developed countries to improve the functioning of their oral and pharyngeal muscles [2]. The main aim of such training programs is to reduce the risk from the elements of dysphagia. In the later stages, augmentative and alternative communication (AAC) strategies were introduced for children with such disorders [3]. These strategies are helpful in providing the expressive and receptive language development modalities to improve the speech and spoken languages. Later Cumley and Swanson produced three case studies based on AAC for children with Developmental Apraxia of Speech (DAS) using speech, signs, and other aids included in child's intervention plan [4].Integral stimulation methods were introduced in the early stages, reported by Strand and Skinder used for articulatory and dysarthria treatment [5, 6].Integral stimulation helps to treat children with motor planning deficit (i.e. developmental apraxia of speech, DAOS) disorders.To improve the speech, language and communication among the children suffering with different disorders needs to train them learn different aspects related with behaviour, emotions, thinking, social, play, learning,reading and writing, and problem solving as shown in Fig. 1.

The communication disorder in children with velocardiofacial syndrome (VCFS) and Down syndrome were studied by Scherer et al. to understand the communication levels of children by using the cognitive and speech and language assessment [7]. A serious differentiation was identified in children with VCFS with respect to receptive and expressive languages, and also with speech sound acquisitions. Pennington et al. tried to understand the effectiveness of speech and language therapy (SLT) and its influence on children while they are communicating with others. Interaction patterns are being tested and measured for the different changes that occur over a period of time [8]. Based on the affects of Down syndrome Kumin conducted a survey to find out the specific factor which influences the speech intelligibility [9]. The children with Down syndrome found to be facing difficulties with voluntary programs, combine different aspects of events, to organize them properly, and to sequence the movements that are required for speech process. Alternative and augmentative communication (AAC) system was suggested by American Speech-Language-Hearing Association (ASHA) to deal with the children suffering with autism [10]. Different strategies, symbols, components, printed words, objects, etc. are defined the AAC system, to improve the communication of the children suffering with autism. Later, Light and Drager extended similar work (AAC) for the young children who suffer with complex communication issues [11, 30]. The usage of AAC helped to apply better tools to solve and improve young children with variety of complex communication issues. A similar research was carried out by Roberts *et al.* to identify the development procedures of language and communication in the children suffering from Down syndrome [12]. This research explained how the language development was established and communication behaviour at the paralinguistic periods, adolescent stages and adult stage.

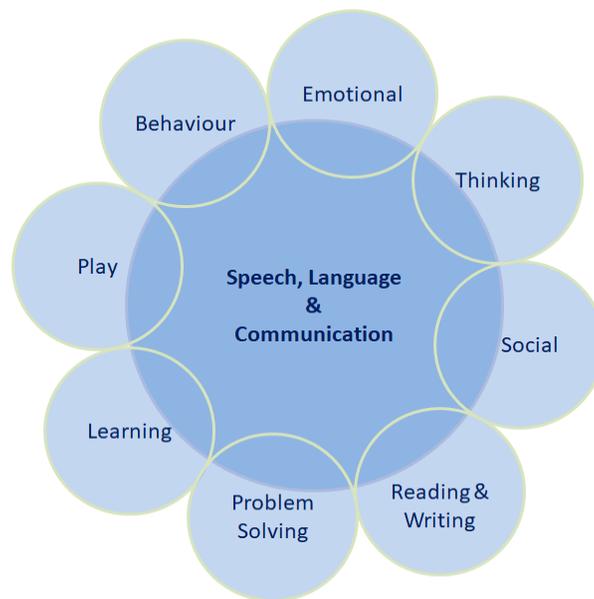

Fig. 1: Areas to be focused to improve the speech, language and communication for children with various disorders

In an attempt to explore the communication deficits of children with autism spectrum disorders (SAD), Paul discussed interventions for improving the communication in two ways as: a) early and prelinguistic communicative behaviours, and b) examining the intervention of children with difficulty for social interaction [13]. The author in this work expressed an efficacy and weakness of behavioural programs and development of communication styles. Computer-presented social stories and different video models were applied to improve the communication of children with autism [14]. This method proved to be very useful to improve rate of social communication and modifications are advised to allow social reinforcement. However, Lang *et al*. suggested the role of parents and the training they need to go through for improving the communication interventions for children with ASD [15]. The training sessions for parents found to be helpful for improving the methods of verbal instructions, role playing, reviving the videos of different intervention sessions. Later on the information and experience of parents along with the information obtained from the clinicians were used to describe the changes observed by following different types of treatments by Thomas-Stonell *et al*. [16]. The observations revealed positive changes across International Classification of Functioning, Disability and Health-Child and Youth version (ICF-CY) frame work domains with respect to participation and personal factors.

A popular communication and speech training program, Picture Exchange Communication System (PECS) was analysed by Flippin *et al*. for the young children with ASD [17].From this work, the meta-analysis synthesis observed a positive gain with respect to communication, but lack of gain was observed in case of speech using PECS [26, 31]. Apart from these maintenance and generalization proved to be a big concern. In an attempt to compare the two effectively proven treatments for children with ASD Ingersoll compared naturalistic behavioural and development with the social pragmatic approaches [18]. Different types of critical differences among these two methods that influence the children with ASD are highlighted in this work and suggested to have much more effective interventions with respect to social-communication. In a different way by adding the linguistic information with music stimuli various principles of pattern perceptions, Lim produced variety of functional speech to help the speech and training programs for children with ASD [19]. In this study, the author tried to compare the effect of training with music, speech and no-training on verbal productions.Both high and low functioning children with ASD improved the speech production due to the training.

Some of the traditional researchers include behavioural aspects to deal with autism disorders by applying scientific teaching principles such as applied behaviour analysis (ABA) [20].Similar attempt was made by Ganz *et* al. to test the AAC interventions by using meta-analysis of single case research studies and later similar attempt was carried out by Grynszpan *et al*. [21, 28]. A selective review was provided in this work on different types of interventions using ABA. Later to improve the speech output for children with ASD, Wan *et al*. suggested auditory-motor mapping training (AMMT) for the purpose of intervention [22]. Apart from these computer-based interventions are also applied by Ramdoss *et al*. for teaching the communication skills to the children suffering from autism [23]. In a generalised way, to address the communication problems with and without ASD, Adams *et al*. tried a random control trial to test the effectiveness of speech and language therapy [24].An attempt to provide the research based interventions for the students suffering with ASD is studied by Koegel *et al*. for the inclusive school setting [25]. Preschoolers with ASD and minimal speech were assigned with naturalistic treatment for spoken languages and communication treatment [27]. Norbury introduced social (pragmatic) communication disorder (SPCD) criteria to relate with ASD [29]. These methods proved to be a dimensional symptom profile which helps to identify range of neuro-development disorders.

## 2. Different Social Skills Between Normal And Disordered Children

The children with different age groups display variety of social skills with respect to their age, after every six months different skills are displayed. This in turn helps the parents and doctors to identify the classification between normal and abnormal kids.

### a) Children with Normal Behaviour

The children at the early age (i.e. between 3 to 6 months) distinguish between the familiar faces and strangers. They display the happiness to see their parents and like to play with others. Sometimes they even respond to the emotions of people and always feel happy to see themselves in mirrors. In case of a child with 12 months old will start pointing the people, responds when called with name, and express shyness or nervousness for strangers. Starts expressing their interests and hate for few things, people, situations, and cries when parents are leaving [32]. Express their interest towards a book, story, games, and sometimes they repeat sounds or the actions to gain attention of people.

In case of children with 18 month or above age, would like to play by passing the things to others and show affection to familiar people with a sense of acknowledgement. They tend to get afraid of strangers and might have temper tantrums. They pretend to feed dolls, cling to caregivers and tries to explore in the presence of parents. In case of 24 month and above age children, they try to copy others (especially elder children), get excited with the group of children, try to be independent, and try to play along with other children [32].

### b) Children Behaviour with Disorders

The children shy temperament will have the eye contact with new people except parents and caregivers. However, children suffering with ASD are not watchful and for referencing also they will not look up to the caregivers. But at the time of anxiety they adhere to the caregiver's or parents. They posses quietness, slow to make new friends, looks down, and gets comfortable slowly when in groups. Most of the children suffering with disorders find it difficult to communicate or share their feelings, enjoyment, achievements and interests. Hardly they will be able to establish or develop new relationships and fails to get acquaintance with peers also. Sometimes they find it difficult to establish the eye-to-eye contact with family members and friends. Lack of emotional feelings is another area of concern for these children along with back to back conversation. Some of the indicators for disorder children are classified and listed in the Table 1 with respect to different domains.

Table 1: Indicators of children with disorders based on different domains

| Language Domain | Behavioural Domain | Cognitive Domain or Play Behaviour |
|---|---|---|
| Pointing the pictures or goods when they are named | Try to stand on tiptoes and kicks the ball when they see on ground | Tries to find the hidden things and likes to play this as a game |
| Remembers the body parts and nearest people | Will try to run when they are left alone | Tries to arrange the shapes, colours and classifies products |
| Speaks small / discrete sentences (in general they contain two to four words) | They walk on the stairs up and down by holding the support or walls | Try to complete the sentences and rhymes |
| Understands simple instructions and get confused with long lines | Tries to climb on the furniture and get down repeatedly | Try to build the towers of blocks, books, etc. |
| Repeated words are used when they speak | Throws the ball repeatedly here and there | Mostly uses one hand as compared to the other |
| Points different things when they see at any place | Tries to copy the circles or straight lines | Two-step instructions are easily followed with happiness |

## c) Children with Autism Spectrum Disorder (ASD)

The behaviour of children with ASD is shown in Fig. 2 with a Triad representing different issues categorised in three broad ways such as i) Impairments in Imagination, ii) Impairments in social interaction and iii) Impairments in social Communication [33].

**i) Impairments in Imagination:** There will be a lot of problem towards thinking and lack of flexibility is seen towards different interests, routine, opinion and rules. Some of the highlights are given below in this category:

- Finds it difficult to understand opinions and feelings of others
- Agitates with routines changes
- Finds it difficult to generalise the information
- They will have some kind of special interest
- Literally takes everything within their reach

**ii) Impairments in social interaction:** There will be a lot of problem towards behaviour and interaction while speaking to others. Some of the highlights are given below in this category:

- They touch others in an inappropriate way
- Find it difficult to understand others and uses nonverbal behaviour (eye contact, face expressions, gestures, etc.)
- Always try to associate with close people
- Do not know different ways of interacting with other people
- Tries to make new friends and relatives with lot of struggle and initiation

**iii) Impairments in social Communication:** There will be a problem to communicate with others in an effective manner. Some of the highlights are given below in this category:

- Repeats the same question
- Will feel difficulty to read the in between lines
- Stick to own lines by ignoring others response

- Makes inappropriate comments which does not suits for the context
- Show less interest to communicate with others
- Communicates only when they need something

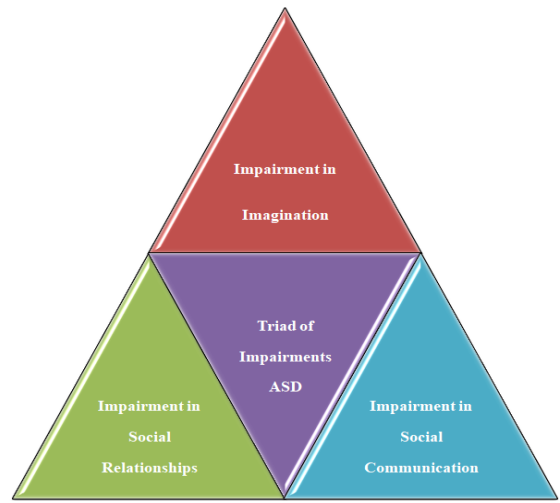

Fig. 2: Triad of Impairments for a child with ASD

Apart from these there are some **_additional difficulties_** seen in these children such as sensory, mental, physical difficulties, etc.Also they cannot bear louder noises, gets hyper for touch, clothes and some kind of external pressures. They tend to have fluctuated moods, anxiety, depressions and aggressions. Also walks on tip-toes, gets distracted easily.

Now-a-day the world is looking into various technological support systems to help the children suffering with variety of disorders, related with speech and communications. For solving such imperative issues artificial intelligence (AI) based technologies are widely used in designing the ROBOTIC TOY's to help the children. These toys are well programmed and intelligent to help the children with disorders to communicate, correct and advice them whenever it is necessary. Joseph Michaelis, a research student from UW-Madison designed a robot which learns how to be a companion and help the kids with thoughtful comments, such as "*... oh, wow I'm scared*" [34]. In the next section, how these robots will map the emotions and are executed to train the robots will be explained. At the same time how they express their emotions with children to help them improve speech and communication will be explained in detail.

## 3. PROPOSED ARCHITECTURE FOR SMART TOY

The *Smart Toy* proposed in this work as shown in Fig. 3 consists of the following modules: a) power supply, b) internet connectivity, c) Monitoring Devices (Sensors/ Cameras), d) Speakers (with voice modulation-sensitive to kid's response), e) database, f) processor, g) prediction module, and h) alert system (to send emails/ SMS's). This smart toy tries to help the children with disorder to improve their communication by helping them to correct the sentences and monitors the child constantly. Different modules are interfaced with different technologies to obtain the assumed results and the detailed explanation is given below:

### a) Power Supply

All the modules in the smart toy are supposed to be supplied with required power supply based on the requirement. This modules must have uninterrupted power supply. Therefore a battery backup is suggested to ensure that there is no power disconnection when the actual power supply is OFF. The modules involved in the *smart toy* are working on 5V, 9 V and 12 V sections based on the circuit design for each module. To avoid any kind of damages due to high voltage fluctuations or short circuit problems, this module is highly protected using a fuse system. This helps to safe guard the *smart toy* to safe for a any kind of unexpected power fluctuations, short circuits and power failures.

### b) Internet Connectivity

Internet connectivity is very much essential to establish a communication between the *smart toy* and parent by sending the information through the emails and SMS's,when unrealistic behaviour the child is observed and when attention of parents is needed.

### c) Runtime Monitoring (Sensors/ Cameras)

The runtime monitoring play a critical role towards observing the child continuously and transmit the data to the database for processing the data to make needful decisions. Apart from that the movement of *smart toy* is required based on the child's movement, therefore, the role of sensors is going to play a crucial role to sense the child's movement and adjust the position of the toy according to the requirement of camera's to capture the child information on continuous basis. The runtime monitoring module focus to capture the following changes by observing a child:

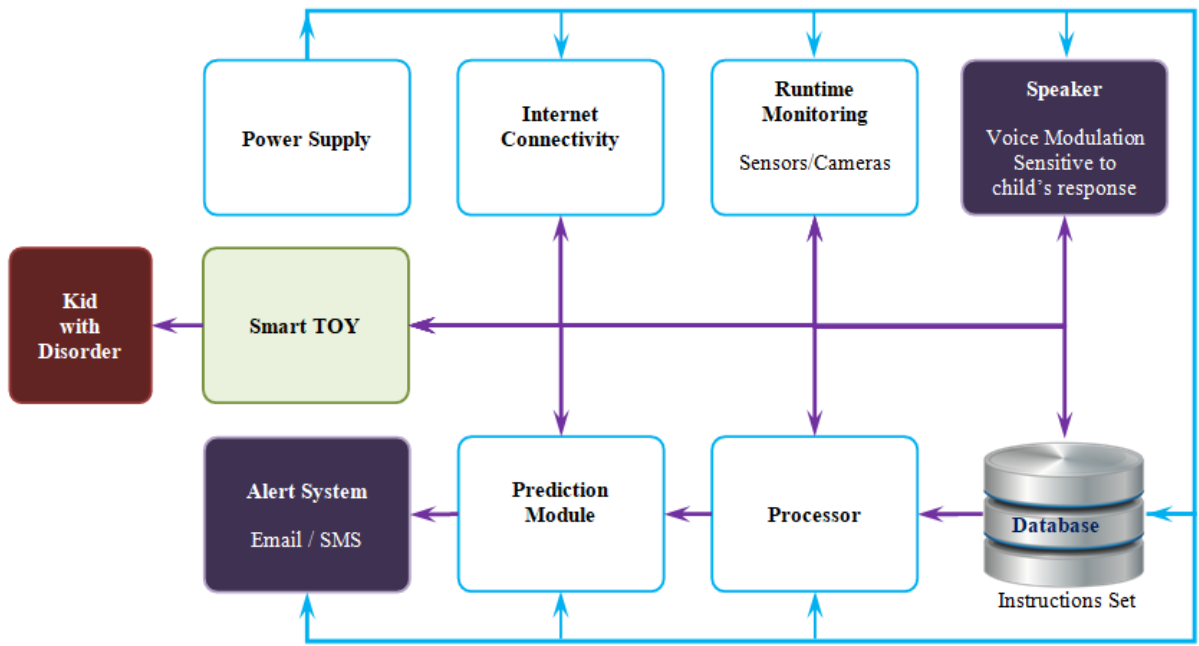

Fig. 3: Proposed Architecture for the children with disorders

**i) Face Expressions and Emotional Changes:** The face expressions are going to play a vital role to understand the emotional changes in a child. The attitude of a child will be revealed in the form of expressions and based on that the *smart toy* needs to make decisions by comparing the stored information in the database. Lot of research in this direction is already in progress by various scientists and some of the image processing techniques are used.

**ii) Walking Behaviour:** The walking behaviour of children with ASD is different as compared to the normal kids and some of the symptoms emotions are also can be read out of the walking behaviour of children.

**iii) Observation of child's voice:** The tone of the kids changes with different emotions and situations. Based on the changes in the voice, the *smart toy* can predict the intentions of the kid. Whenever the child speaks a wrong word out of any emotion, the *smart toy* will take it as an input and reply with suitable voice outputs.

### d) Speakers

The Speakers arranged with the *smart toy* plays a key role to establish the communication with the child and needs to have sensitive voice modulation technique to respond with kid's emotions and behaviour. Here, voice modulation is highlighted because sometimes the child may like the primary voice adjusted at the output of the speakers. In such scenarios, the variation of tones may help the child to get some kind of happiness. For example, if the instructions are given in a male voice are irritating the child means, the speakers can deliver a female voice.

With the help of advanced technologies available in the market, the family member's voice also can be recorded / stored in the database and deliver the needful instructions in their tone. Such facility will help the *smart toy* to convince the child with ASD to some extent. Therefore a sensitive voice modulating module is going to help the *smart toy* to get more attention of the disordered kids to communicate and correct their mistakes while speaking to others.

### e) Database

The database needs to provide assistance to various modules in the proposed architecture by saving the large amount of the data that is captured by runtime monitoring module and allow the same data to be transferred to different modules to make a proper prediction to deliver a resultant output at the speakers.The database needs to have a process of constant learning, monitoring inputs, prediction results / summaries, training data, error correction inputs/ outputs, time bounding schedules, and alerts. Different components involved with the database are shown in Fig. 4.

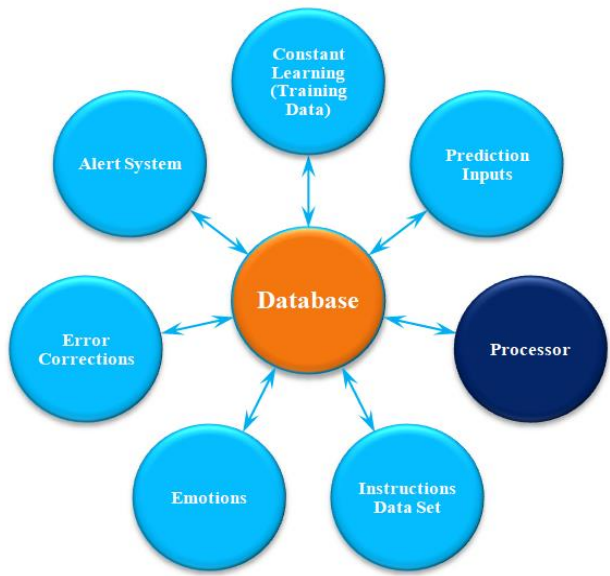

Fig. 4**:** Database and Interacting Components

**f) Processor**

In this architecture, the role of a processor is to provide the needful controls and to operate on a priority based instructions. The processor will control all the inputs and outputs to be delivered at different levels of operations to improve the quality and performance of the *smart toy*.

**g) Prediction Module**

This module is going to play a key role to assess the inputs from different types of monitoring devices and compare them with existing data to deliver the best suitable outputs at the end. This module consists of four blocks: i) Input, ii) Pre Processing, iii) Processing and iv) Output block as shown in Fig. 5. The processing block again sub divided into three blocks as i) feature-extraction, ii) feature-classification and iii) feature-model computation.

**i) Input**

In the first stage, different facial images are detected from the monitoring devices as the input images. These images will be having a set of sequence to identify the facial components such as eyes, nose, etc. At the same time some of the voice inputs are also going to be applied at the input for the processing and to improve the communication among the disordered children.

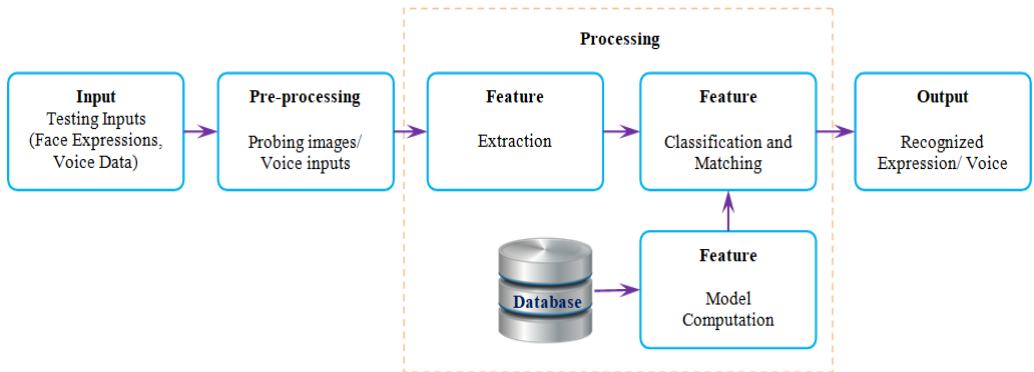

Fig. 5: Children's Emotions Prediction Module

**ii) Pre-processing**

The pictures collected from a camera / video will have many problems, which needs to be addressed properly. Some of the areas where the pre-processing can improve and correct the images using different types of descriptor techniques such as local binary, spectra, basis space and polygon shape descriptors. These descriptors will help to improve some of the features like illumination corrections, blur and focus corrections, filtering and noise removal, thresholding, edge enhancements, morphology, segmentation, region processing and filters, point processing, math and statistical processing and colour space conversions [36].

At the time of pre-processing some of the artefacts of the images must be corrected before the feature measurement and analysis. Such corrections include sensors based, lighting based, noise based, geometric and colour based corrections. Apart from that for specific feature measurements, some kinds of enhancements are used to optimise instead of fixing the problem. Some of the image enhancement techniques are scale-space pyramids, illumination changes, and blur and focus enhancements [36]. Similarly the audio pre-processing also will be taking place in this section to understand the intentions of the child behaviour. The neural networks based classification techniques are applied for the classification of voice and images in this module. Also back propagation algorithm was used to predict the suitable values by comparing the original images with the properties stored in the database. Audio processing in this module is based on the time frequency representation, frequency weighting and scaling techniques [37, 38].

**iii) Processing**

The processing module contains three blocks and works along side with the database to compute various feature classification modules.

- **Feature-Extraction:** In this block the facial emotion recognition (FER) and voice recognition processes will be taking place out of many sample images as shown in Fig. 6. All the input images will be extracted for face detection and then the similar faces detected will be applied for the landmark detection.

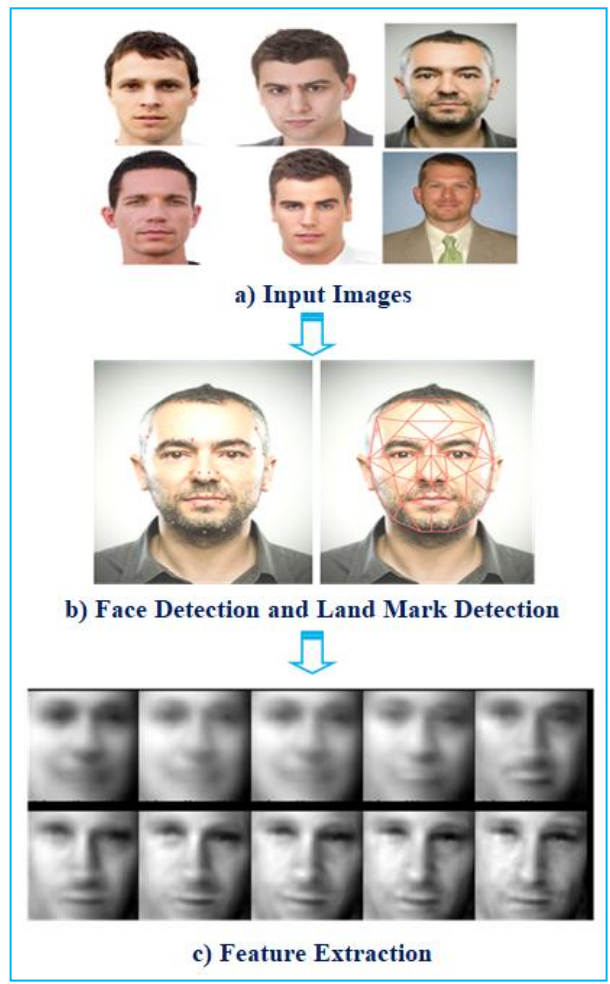

Fig. 6. A conventional procedure for FER approaches to detect face expressions

- **Feature-Classification:** In this block the image are will be processed for landmark classification as shown in Fig. 7 for understanding different emotions from the facial readings.

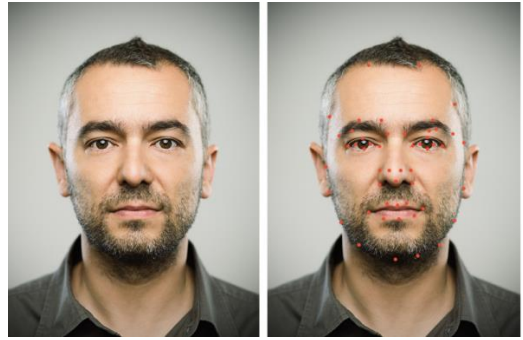

Fig. 7. An image processed for landmark classification.

- **Feature-Model Computation:** The determined features of the facial expressions are then compared with the database content to identify the emotions and behaviour. This type of study helps to understand the changes in child requirements and according to that the *smart toy* can produce the communication outputs.

### iv) Output Block

The output block will have all the filtered information about the emotions of the child and based on which the decisions for generating voice commands will be given to the *smart toy*. These commands will be processed by again by observing the response of the child. For example, if the child is not impressed with male voice, the *smart toy* automatically changes it voice to produce a female voice version. Sometimes it will also produce the corrections for the child if the pronunciation of the words is not correct and tries to improve the communication standards also. To carry out such events the *smart toy* needs to undergo for the constant training process with the familiar sounds (of family members, scenarios and remedies for health related problems). This training part will be taken care by the neural network based systems in the proposed architecture.

### h) Alert System

The alerts are produced in the extreme conditions when the monitoring section identifies such events through the emails or SMS's. The prediction module will provide the inputs for the alert system by comparing multiple events with the already stored instructions in the database. For example, the children suffering with disorders tend to hit their head to the walls or repeatedly they tend to beat their legs on the floor, etc. Such events will generate the alerts for the parents to take immediate action.

## 4. PROPOSED CHECKER COMPONENT AND POLICIES RULES

The proposed checker component is dealing with different types of interdisciplinary datasets. The monitoring devices included with cameras, sensors and voice detectors will provide the inputs for the prediction module. This arrangement was tested in real-time environment using embedded system applications. In this paper, a checker component was proposed to produce the alerts in the form of emails and SMS's as shown in Fig. 8 with the flow process of the events.

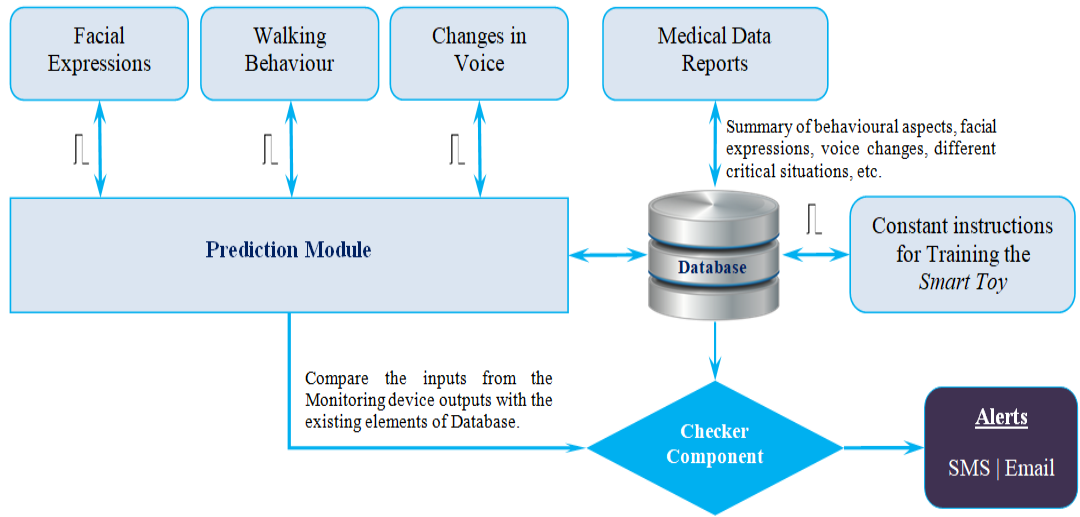

Fig. 8**.** Runtime Checker Component Design for Alert System

Some of the policies are framed to understand the behaviour or the children and to establish the efficient communication between the disordered child and the *smart toy* are given below:

☞ *Face Expressions and Emotional Changes:*The face expressions are extracted from the cameras and based on these expressions and the processed data after comparison the decision must be made to deliver the next action.

☞ *Walking Behaviour:*The walking behaviour of the kids will be observed and based on the walking behaviour the mood (happy, angry, sad, etc.) of the children is predicted.

☞ *Observation of child's voice:* The tone of the kids will be observed carefully to understand whether the kid is speaking with happiness, fear, or any other intention.

☞ *Response for the Smart Toy:* This is important to understand what kind of voice made by the smart toy is making the child happy and based on the behaviour of the child for a particular voice (female or male) version will be produced by the *smart toy*.

☞ *Constant training for Smart Toy:* The smart toys are supposed to understand and learn the new events and developments of the child environment. Such as a new guest visited the home, or a new situation arises at home, a new member joined the home, etc. Such observations and inputs of concerned event must be trained to the *smart toy*.

Based on the above events the policies are defined for the *smart toy* is listed below:

The smart toy (*S*) needs to address the facial expressions (*FE*) and emotional changes (*E*) as defined by the instructions of the experts are formalized here:

$$Policy1 \triangleq$$

$$\begin{pmatrix} fin(SmartToy) \wedge \\ Face\ Expressions\ (FE,\ Module) \wedge \\ Emotions\ (E, Module) \wedge \\ \wedge_{i=Environment}^{t=Expressions,Emotions}\ done\ (FE, E, Submit) \end{pmatrix} \mapsto \left( Authorize^+(FE, E, Approve\ for\ S) \right)$$

The smart toy ($S$) needs to look after the walking behaviour (WB) from different datasets (DB) and observations (O) are formalized here:

$$Policy2 \triangleq$$

$$\begin{pmatrix} fin(SmartToy) \wedge \\ WalkingBehaviour\ (WB,\ Module) \wedge \\ Database\ (DB, Module) \wedge \\ \wedge_{i=Environment}^{t=Walking\ Behaviour}\ done\ (WB, DB, Submit) \end{pmatrix} \mapsto \left( Authorize^+(WB, DB, Approve\ for\ S) \right)$$

Now to understand the child voice ($V$) and the modulation in voice (M) with different times ($T$) and scenarios are drafted for the smart toy operations.

$$Policy3 \triangleq$$

$$\begin{pmatrix} fin(Energy, Maintain : Essential) \wedge \\ Voice\ (V,\ Module) \wedge \\ Modulation\ (M, Module) \wedge \\ \wedge_{i=0}^{i=Matching}\ done\ (V, M, Successful communication) \end{pmatrix} \mapsto \left( Authorize^+(V, M, Communication) \right)$$

Here the child's response for a particular voice will be tested. If the emotions and expressions of the child are not convincing with male voice (MV) means the smart toy needs to change that to a female voice (FV), so as to see child with happy behaviour.

$$Policy4 \triangleq$$

$$\begin{pmatrix} fin(ChildBehaviour) \wedge \\ MaleVoice\ (MV,\ Module) \wedge \\ FemaleVoice\ (FV, Module) \wedge \\ \wedge_{i=Environment}^{t=MaleVoice,FemaleVoice}\ done\ (MV, FV, Submit) \end{pmatrix} \mapsto \left( Authorize^+(MV, FV, VoiceModulation) \right)$$

Constant training is needed for the smart toy with the changes in the environment and needs to learn (L) new voices and train the smart toy (TR) to establish the communication changes at the same with the kids.

$$Policy5 \triangleq$$

$$\begin{pmatrix} fin(Learning, Training : SmartToyUpdates) \wedge \\ Learning\ (L,\ Module) \wedge \\ Training\ (TR, Module) \wedge \\ \wedge_{i=0}^{i=Fulfilled}\ done\ (L, TR, UpdatedSmartToy) \end{pmatrix} \mapsto \left( Authorize^+(L, TR, SmartToyUpdated) \right)$$

The policy rules defined above will help the smart toys to deliver a meaningful communication for the child with disorders by establishing a middleware between hardware and software used for the design of smart toy. The proposed *smart toy* will be able to direct the children suffering with communication disorders to get instructions and suggestions as shown in Fig. 9.

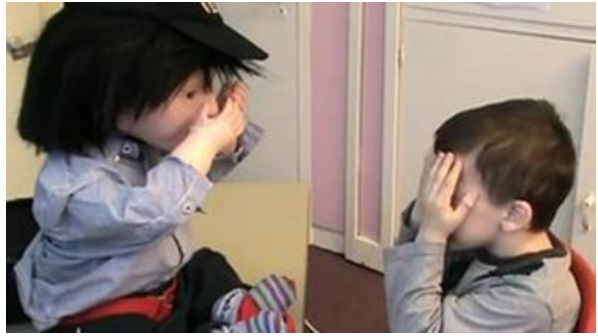

Fig. 9. A smart toy instructing the child with disorder to follow the instructions based on the predictions

## 5. CONCLUSIONS

The proposed smart toy proved to be a better companion for the children suffering with communication disorders, since it is well equipped with constant learning and training properties. The communication between various monitoring devices and prediction components are processed at a faster rate so as to make an immediate decision is delivered by the smart toy to establish a meaningful communication. Above and all, the proposed system using neural networks with the help of back propagation algorithm generates quickest alerts using emails and SMS's to the parents at a faster speed so that serious damages can be avoided to the children suffering with communication disorders. This work speaks the applications of fuzzy neural networks which help to enhance the intelligence of the proposed application which helps both parents and doctors to detect the children suffering with autism. The policies proposed in this paper will help the smart toy to work more accurately and efficiently.

**Author's Biography**

Amr Jadi was born in Medina of Saudi Arabia, onJuly 9, 1985. The author is involved in various scientific research activities of the Department of Computer Science and Information, University of Hail.Dr. Jadi received honorary degrees from De Montfort University and Masters Degree from Bradford University, UK. The author is specialized in Computer Sciences with an area interest in Early warning systems, Risk management and Critical Systems. Presently the author is also involved in various development activities within the University of Hail and abroad as a consultant.

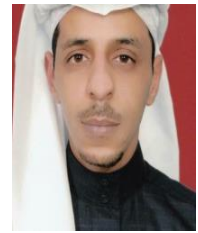